\documentclass[amsfonts,amsmath,amssymb,nofootinbib]{revtex4}

\usepackage{ifpdf}

\ifpdf
    \usepackage[pdftex,colorlinks=true,plainpages=false,pdfpagelabels]{hyperref}
    \usepackage[pdftex]{color}
    \usepackage[pdftex]{graphicx}
    \hypersetup{pdfauthor = {John Kehayias and Stefano Profumo}}
    \hypersetup{pdftitle = {Semi-Analytic Calculation of the Gravitational Wave Signal From the Electroweak Phase Transition for General Quartic Scalar Effective Potentials}}
    \hypersetup{colorlinks = true}
\else
    \usepackage[colorlinks=true]{hyperref}
    \usepackage{graphicx}
\fi

\numberwithin{equation}{section}

\begin{document}

\title{Semi-Analytic Calculation of the Gravitational Wave Signal From the Electroweak Phase Transition for General Quartic Scalar Effective Potentials}

\author{John Kehayias}
\email{kehayias@physics.ucsc.edu}
\affiliation{Santa Cruz Institute for Particle Physics and Department of Physics,\\ University of California, Santa Cruz, CA 95064, USA}

\author{Stefano Profumo}
\email{profumo@scipp.ucsc.edu}
\affiliation{Santa Cruz Institute for Particle Physics and Department of Physics,\\ University of California, Santa Cruz, CA 95064, USA}

\date{\today}

\begin{abstract}
\noindent Upcoming gravitational wave (GW) detectors might detect a stochastic background of GWs potentially arising from many possible sources, including bubble collisions from a strongly first-order electroweak phase transition.  We investigate whether it is possible to connect, via a semi-analytical approximation to the tunneling rate of scalar fields with quartic potentials, the GW signal through detonations with the parameters entering the potential that drives the electroweak phase transition.  To this end, we consider a finite temperature effective potential similar in form to the Higgs potential in the Standard Model (SM).  In the context of a semi-analytic approximation to the three dimensional Euclidean action, we derive a general approximate form for the tunneling temperature and the relevant GW parameters.  We explore the GW signal across the parameter space describing the potential which drives the phase transition.  We comment on the potential detectability of a GW signal with future experiments, and physical relevance of the associated potential parameters in the context of theories which have effective potentials similar in form to that of the SM.  In particular we consider singlet, triplet, higher dimensional operators, and top-flavor extensions to the Higgs sector of the SM.  We find that the addition of a temperature independent cubic term in the potential, arising from a gauge singlet for instance, can greatly enhance the GW power.  The other parameters have milder, but potentially noticeable, effects.
\end{abstract}

\maketitle


\section{Introduction}

Gravitational wave (GW) interferometers currently taking data, and under development or construction, have as a possible target a stochastic background of GWs of cosmological origin (for reviews, see\cite{thorne,allen:1996vm,Maggiore:1999vm,Maggiore:2000gv}, for recent results, see \cite{ligonature}). Since GWs propagate freely through the Universe after being produced, their detection provides a powerful diagnostics for the physics of the Early Universe. Several mechanisms that might generate such GWs have been discussed, including quantum fluctuations during or shortly after inflation \cite{Turner:1990rc,hogan,Witten:1984rs}, cosmic strings \cite{Steinhardt:1981ct}, cosmological magnetic fields \cite{Gyulassy:1983rq,VanHove:1983rk,Enqvist:1991xw,Huet:1992ex}, plasma turbulence \cite{landau,Abney:1993ku,weinberggr,Turner:1992tz} and bubble wall collisions during first-order phase transitions \cite{Kosowsky:1992rz,Kosowsky:1991ua,Kosowsky:1992vn,Coleman:1977py,Callan:1977pt,Kamionkowski:1992dc}. 

Specifically, the space interferometer LISA, expected to fly, or to be nearing completion, over the next decade will have a sensitivity peak for GWs with a frequency between $10^{-4}$ and $1$ Hz \cite{Maggiore:2000gv}. Quite fortunately, this range corresponds to the frequency range expected today, after redshifting, from GWs produced at a temperature $T \sim 100\ {\rm GeV}\sim E_{\rm EW}$, where the latter symbol indicates the electroweak scale. Intriguingly enough, the production of GWs at $T\sim E_{\rm EW}$ is indeed expected if the electroweak phase transition is strongly first order. In turn, this is a necessary condition for the success of scenarios where the baryon asymmetry of the Universe is produced at the electroweak phase transition (electroweak baryogenesis, EWB; for a review, see \cite{Riotto:1999yt}).  However, there is a generic tension between the requirement of a small bubble wall velocity needed in the context of electroweak baryogenesis models (see e.g.~\cite{bauref1,bauref2,bauref3}) and the super-sonic bubble velocity values (detonation) under investigation here.  Depending on the sources of net baryon number, super-sonic bubble wall velocities might be compatible with a baryon asymmetry produced at the EWPT, although this is generally not the case.  We are at the dawn of the exciting period of exploring the electroweak scale with the possibility to connect such diverse experimental endeavors as GW detection and the Large Hadron Collider.  These grand experimental enterprises may also help answer the question of how the baryon asymmetry arose in the Early Universe, which is all the more exciting.


In its minimal setup, the strength of the electroweak phase transition in the Standard Model only depends only on the mass of the Higgs bosons (for a review, see \cite{tempreview}).  For the Higgs mass range compatible with searches at LEP-II, non-perturbative lattice computations indicate that there is no phase transition at all, but rather a smooth crossover \cite{Kajantie:1996mn}.  No GW production is thus expected at the EW phase transition, if the electroweak sector corresponds to the minimal Standard Model.  In addition, if this is the case, EWB cannot be the mechanism underlying the production of the baryon asymmetry in the Universe.

Fortunately, the strength of the EW phase transition is actually strongly model-dependent and parameter-dependent (for a review, see \cite{Rubakov:1996vz}), and numerous extensions of the Standard Model do predict a strongly first order EW phase transition. If this is further extended with the other necessary ingredients for EWB, the EW scale might indeed be responsible for generating the observed baryon-antibaryon asymmetry. 

If the EW phase transition is strongly first order, the universe finds itself trapped, at $T\lesssim E_{\rm EW}$, in a metastable EW unbroken phase where the vacuum state of the universe is a false vacuum (i.e.~it is not the lowest energy state) as the universe cools down and its temperature decreases.  A potential energy barrier exists between the false and the true (lowest energy) electroweak symmetry breaking vacuum.  Quantum mechanical tunneling produces bubbles of true vacuum (broken EW phase), which then expand, collide, and combine to fill the universe with the true vacuum.  GWs can be abundantly produced at the EW phase transition, primarily through bubble collisions \cite{Witten:1984rs,hogan,Turner:1990rc,Kosowsky:1992rz,Kosowsky:1991ua,Turner:1992tz,Kosowsky:1992vn,Kamionkowski:1993fg}.  This mechanism has been extensively investigated, analytically and numerically, and in relation to the effective potential that drives the transition itself.

In particular, the possibility of a GW signal from the EW phase transition has attracted, not surprisingly, a great deal of attention.  In view of the above mentioned large model-dependence, though, most of the existing literature has been devoted to either special cases and particular corners of parameter space (such as the light stop scenario in the context of the minimal supersymmetric extension of the Standard Model, MSSM; for a review, see \cite{Quiros:2000wk}), and/or to accurate but numerical studies only \cite{Ashoorioon:2009nf,gwbubble}.  Alternatively, model independent results on the detectability of a GW signal from a strongly first order phase transition have been derived assuming a few relevant parameters for the GW dynamics could be computed from the effective potential that drives the particular phase transition under consideration (for recent related work, see e.g.~\cite{Grojean:2006bp,Caprini:2007xq,Kahniashvili:2008pf}).

The scope of the present study is to use semi-analytical results of the tunneling rate of scalar fields with quartic potentials \cite{se3approx} to predict {\em analytically} the strength of the EW phase transition and the GW signal in extensions of the Standard Model EW sector that can be characterized with the dynamics of a single order parameter, a scalar field $\phi$.  These models include simple generalizations of the SM effective potential, in appropriate dynamical regimes.  The main result of this paper is a closed analytic formula for two parameters describing the GW signal as a function of parameters appearing in the effective potential of the scalar field driving the EW phase transition.


The rest of this paper is organized as follows: in Section \ref{sec:backgrnd} we summarize the relevant physics and definitions of the quantities we will be studying, in particular an approximation for the three dimensional Euclidean action, which is the basis for computing the finite temperature tunneling rate.  Then, in Section \ref{sec:approx}, we derive an approximation for the tunneling temperature and present exact and approximate formulas for the GW parameters for any effective potential similar in form to the SM case.  Following that, in Section \ref{sec:param} we constrain some parameters in order to reproduce the usual electroweak symmetry breaking pattern, and plot the effect of the various parameters in the potential on the parameters driving the GW signal.  We examine the physical relevance of the parameter space and the detectability of the GW when varying various parameters beyond their SM values.  In Section \ref{sec:models} we describe specific examples where our formalism can be applied in certain regimes, including $SU(2)$ triplet and singlet extensions to the Higgs sector, top-flavor models, and models encompassing higher dimensional operators. 

\section{Background}\label{sec:backgrnd}

In the context of a strongly first-order phase transition in the early universe, the basic problem of evaluating the GW signal amounts to calculating the tunneling rate (i.e.~the decay probability) from the false vacuum state to the true vacuum state -- in other words, the bubble nucleation rate per unit time per unit volume (for a review of tunneling in a finite temperature quantum field theory, see \cite{tempreview}). This is given in general by the expression:
%
%
%
\begin{equation}
\frac{\Gamma}{V} \sim A(T)e^{-S_{E3}/T}
\end{equation}
where the factor $A(T) \sim T^4$ and $S_{E3}$ is the three dimensional\footnote{The symmetry for finite temperature bubble nucleation is $O(3)$, not $O(4)$ \cite{linde1,linde2}.} Euclidean action
\begin{equation}
S_{E3} = \int\textrm{d}^3x\left[\frac{1}{2}\left(\vec{\nabla}\phi\right)^2 + V\left(\phi,T\right)\right].
\end{equation}
Typically $S_{E3}$ is not calculated analytically, but semi-analytic expressions exist (\cite{se3approx}), which provide the basis for the rest of our analysis.  We summarize these results below.

As we point out in Sec.~\ref{sec:approx}, it is possible to derive an expression for the temperature at which tunneling is probable and the universe undergoes a phase transition.  Now that regions of the universe can tunnel to the true vacuum, it is possible to produce large bubbles of true vacuum which expand, collide, and combine into larger bubbles.  Expanding bubbles gain wall velocity and energy, but spherical symmetry does not allow any energy to be directly transferred into GWs.  However, when more than two bubbles collide, this symmetry is broken and energy can be released into GWs\footnote{Section IIB of \cite{Kosowsky:1991ua} has a detailed discussion of the symmetry of two colliding bubbles.}.  Energy released into the universe can also be transferred to GWs through turbulence, but we neglect this sub-dominant contribution to the spectrum (for recent analyses, see \cite{Kosowsky:2001xp,Gogoberidze:2007an,Caprini:2009yp}).

Two quantities determine the GW spectrum when the phase transition proceeds through detonation (the bubble wall velocity is faster than the speed of sound in the plasma).  $\alpha$ measures the energy density change by transitioning from the false to true vacuum, and $\beta$ is the bubble nucleation rate per unit volume.  The actual parametrization of the GW spectrum is summarized below.

\subsection{The Three Dimensional Euclidean Action}\label{sec:se3}
The tunneling process is calculated from the three dimensional Euclidean action.  Typically this is done numerically, however, there is a general, semi-analytic, approximate solution for quartic potentials of a single scalar field \cite{se3approx}.  This is the key starting point for our calculations.

Consider the potential for a scalar field $\phi$ of the form
\begin{equation}
V(\phi) = \lambda\phi^4 - a\phi^3 + b\phi^2 + c\phi + d
\end{equation}
with $\lambda > 0$ to have the potential bounded from below, $b > 0$ to have $\phi = 0$ a minimum, and $a > 0$ for the minimum to be for positive $\phi$.  Without loss of generality, the false vacuum can be placed at the origin ($\phi = 0, V = 0$), and so $d = c = 0$.  Here the above coefficients are typically not the same as in the zero temperature potential of the same theory, and may have temperature dependence.  They are also all greater than zero.  The three dimensional Euclidean action is approximately given by
\begin{equation}\label{eq:se3}
S_{E3} = \frac{\pi a}{\lambda^{3/2}}\frac{8\sqrt{2}}{81}(2-\delta)^{-2}\sqrt{\delta/2}\left(\beta_1\delta + \beta_2\delta^2 +\beta_3\delta^3\right)
\end{equation}
where $\delta \equiv 8\lambda b/a^2, \beta_1 = 8.2938, \beta_2 = -5.5330,
\textrm{and }\beta_3 = 0.8180$.  These parameters are the result of a numerical fit in the semi-analytic study\footnote{The absolute errors of the reduced action divided by the thin wall limit action are bounded to be less than $0.033$ -- see \cite{se3approx} for the full details.} \cite{se3approx}.

\subsection{Gravitational Wave Parameters}\label{sec:gwp}
We are interested in the gravitational wave signal from electroweak symmetry breaking, which is mostly produced through bubble (as in which vacuum state) nucleation.  The first part of calculating the GW parameters is to determine at what temperature bubble nucleation will be an energetically favored process.

Assuming temperatures of $\mathcal{O}(100\textrm{ GeV})$, the probability of a single bubble to be nucleated in a horizon volume to be $\sim \mathcal{O}(1)$ is well approximated\footnote{The exponential factor of the tunneling probability ensures that this approximation is valid for a broad range of temperature or energy scale.} in the Early Universe by
\begin{equation}
S_3(T_t)/T_t \sim 140
\end{equation}
where $T_t$ is the tunneling temperature (see e.g.~\cite{tempreview}).  This temperature will be between the critical temperature $T_c$, where there is a degenerate minimum with $\phi \neq 0$ (the only minimum at high $T$ is at $\phi = 0$), and the destabilization temperature $T_{dest}$, where the minimum at $\phi = 0$ is a local maximum:
\begin{equation}
T_{dest} \leq T_t < T_c.
\end{equation}
An approximation is derived for $T_t$ in Section \ref{sec:approx}.

The vacuum energy (latent heat) density in this process is given by the standard statistical mechanics expression
\begin{equation}
\epsilon_t = -V(v(T),T) + T\frac{\textrm{d}}{\textrm{d}T}V(v(T),T)\bigg|_{T_t},
\end{equation}
and the ratio between this and the radiation energy density is
\begin{equation}
\alpha = \frac{30\epsilon_t}{\pi^2g_tT_t^4},
\end{equation}
where $g_t$ is the number of relativistic degrees of freedom at $T_t$.  In a radiation dominated universe the parameter $\beta$ is given by
\begin{equation}
\frac{\beta}{H_t} = T_t\frac{\textrm{d}(S_3(T)/T)}{\textrm{d}T}\bigg|_{T_t}.
\end{equation}
$\alpha$ and $\beta/H_t$ parametrize the GW spectrum \cite{Kamionkowski:1993fg}, which is defined below, following \cite{gwbubble}.

We consider only a phase transition proceeding through detonation -- the bubble wall velocity is faster than the speed of sound in the plasma.  This also ensures that the thin wall approximation is valid, which was used in \cite{gwbubble}.  The wall velocity is \cite{Steinhardt:1981ct}
\begin{equation}
v_b(\alpha) = \frac{\frac{1}{\sqrt{3}} + \sqrt{\alpha^2 + \frac{2}{3}\alpha}}{1 + \alpha},
\end{equation}
which increases with $\alpha$, starting at the speed of sound in the plasma ($1/\sqrt{3}$) up to the speed of light.  However, this may not always be a good assumption, depending on the exact theory for the electroweak phase transition; particle scattering with the bubble wall will affect $v_b$.  For instance, \cite{vb} analyzed the MSSM stop contribution to the friction of the bubble wall in the plasma.  This can greatly decrease $v_b$, down to about $0.05$.  On the other hand, one still expects some scaling with $\alpha$ (which can be quite large and we are not restricting ourselves to just the MSSM), which could counteract these effects.  Since we do not make any assumptions on the underlying theory, it is difficult to say what value of $v_b$ will ultimately take, and we use the above equation for what follows.  We are also restricting ourselves to the case of detonation, so we must have $v_b$ greater than $1/\sqrt{3}$.  This is a non-trivial calculation from the effective potential and bubble dynamics, so for the purposes of this work we must assume that this holds.  For a particular model, however, one must check that $v_b$ is sufficiently large to apply the formulas we will present for the GW spectrum.

How much of the vacuum energy is transferred to the bulk, rather than reheating the bubble, is given by an efficiency factor, again considering only the case of detonations and the same caveats above:
\begin{equation}
\kappa(\alpha) = \frac{1}{1 + 0.715\alpha}\left(0.715\alpha + \frac{4}{27}\sqrt{\frac{3\alpha}{2}}\right)
\end{equation}
For the bubble collision contribution the spectrum is parametrized (close to the peak frequency) as
\begin{equation}
\Omega_{\mathrm{Col}*}(f_{\mathrm{Col}*}) = \widetilde{\Omega}_{\mathrm{Col}*}\frac{(a + b)\tilde{f}_{\mathrm{Col}*}^bf_{\mathrm{Col}*}^a}{b\tilde{f}_{\mathrm{Col}*}^{(a + b)} + af_{\mathrm{Col}*}^{(a + b)}}
\end{equation}
with the peak frequency $\tilde{f}_{\mathrm{Col}*}$ and peak amplitude $\widetilde{\Omega}_{\mathrm{Col}*}$ (as functions of the wall velocity).  The exponents are in the range $a \in [2.66, 2.82]$ and $b \in [0.90, 1.19]$ with the case of large wall velocity, $v_b \approx 1$, having $a \approx 2.8$ and $b \approx 1.0$ (see the numerical analysis of \cite{gwbubble} for details).  The spectrum observed today is found by redshifting:
\begin{align}
\tilde{f_{\mathrm{Col}}} &= 16.5\times10^{-3}\textrm{ mHz}\left(\frac{\tilde{f}_{\mathrm{Col}*}}{\beta}\right)\left(\frac{\beta}{H_t}\right)\left(\frac{T_t}{100\textrm{ GeV}}\right)\left(\frac{g_t}{100}\right)^{1/6}\label{eq:gws1}\\
h^2\widetilde{\Omega}_\mathrm{Col} &= 1.67\times10^{-5}\widetilde{\Omega}_{\mathrm{Col}*}\left(\frac{100}{g_t}\right)^{1/3}\notag\\
&= 1.67\times10^{-5}\tilde{\Delta}\kappa^2\left(\frac{H_t}{\beta}\right)^2\left(\frac{\alpha}{\alpha + 1}\right)^2\left(\frac{100}{g_t}\right)^{1/3}\label{eq:gws2}
\end{align}
with the functions $\tilde{f}_{\mathrm{Col}*}/\beta$ and $\tilde{\Delta}$ approximately \cite{gwbubble}
\begin{align}
\tilde{\Delta} &= \frac{0.11v_b^3}{0.42 + v_b^2}\\
\tilde{f}_{\mathrm{Col}*}/\beta &= \frac{0.62}{1.8 - 0.1v_b + v_b^2}.
\end{align}
Although we use the results of a numerical analysis for the GW spectrum, there has also been recent analytic work on this subject (e.g.~\cite{Caprini:2007xq,Caprini:2009fx}).

To summarize, the three dimensional Euclidean action is used to calculate the tunneling rate to the true, symmetry breaking, minimum of the potential.  An approximation for quartic potentials was given in \cite{se3approx}, which we will employ in this study.  The action is used to directly calculate the tunneling temperature, where this process can produce appreciable amounts of gravitational waves.  Two parameters are calculated from the potential to give the spectrum observed today.

\section{$T_t, \alpha$, and $\beta/H_t$ from a Generic SM-Like Potential}\label{sec:approx}

The potential we consider is a generic, quartic potential modeled after the SM Higgs's effective potential at high temperature.  What follows is a quick overview of the relevant terms and quantities in the SM.  The one loop, finite temperature, effective potential due to the gauge bosons and top quark, expanded at high temperature, is
\begin{equation}
V(\phi,T) = D(T^2-T_{0 SM}^2)\phi^2 - ET\phi^3 + \frac{\lambda_{SM}(T)}{4}\phi^4
\end{equation}
with the coefficients being
\begin{align}
D &= \frac{2m_W^2 + m_Z^2 + 2m_t^2}{8v^2} \approx 0.17\\
E &= \frac{2m_W^3 + m_Z^3}{4\pi v^3} \approx 9.6\times10^{-3}\\
T_{0 SM}^2 &= \frac{m_h^2 - 8Bv^2}{4D} \approx (238.6~\mathrm{GeV})^2\\
B &= \frac{3}{64\pi^2 v^4}(2m_W^4 + m_Z^4 - 4m_t^4) \approx -4.6\times10^{-3}\\
\lambda_{SM}(T) = \lambda_{SM} - \frac{3}{16\pi^2v^4}&\bigg(2m_W^4\log\frac{m_W^2}{A_BT^2} + m_Z^4\log\frac{m_Z^2}{A_BT^2} - 4m_t^4\log\frac{m_t^2}{A_FT^2}\bigg)
\end{align}
where $\log A_B = \log a_b - 3/2, \log A_F = \log a_f - 3/2, a_b = 16\pi^2\exp(3/2 - 2\gamma_E), a_F = \pi^2\exp(3/2 - 2\gamma_E),$ and $\gamma_E \approx 0.5772$ is the Euler-Masccheroni constant \cite{gammae} (for details, see e.g.~\cite{tempreview}).  All the masses (Higgs, W, Z, and top) above refer to the usual zero temperature values, and $v \approx 246$ GeV is the Higgs vacuum expectation value (vev).  The high temperature approximation comes from expanding the thermal bosonic and fermionic functions, $J_{\mathrm{B,F}}$, appearing as the thermal contribution to the one-loop effective potential:
\begin{equation}
\frac{T^4}{2\pi^2}J_{\mathrm{B}}\left[m^2(\phi)/T^2\right],\qquad
\frac{-2\lambda T^4}{2\pi^2}J_{\mathrm{F}}\left[m^2(\phi)/T^2\right],
\end{equation}
where $m$ is the field-dependent mass of the boson or fermion and we are working in units where $k = 1 = \hbar = c$ (so temperature is measured in energy units).  The thermal functions are given by
\begin{equation}
J_{\mathrm{B,F}}\left[m^2\beta^2\right] = \int_0^\infty\mathrm{d}x x^2\log\left[1 \mp e^{-\sqrt{x^2 + m^2\beta^2}}\right],
\end{equation}
and have high temperature expansions
\begin{align}
J_\mathrm{B}(m^2/T^2) &\approx -\frac{\pi^4}{45} + \frac{\pi^2}{12}\frac{m^2}{T^2} - \frac{\pi}{6}\left(\frac{m^2}{T^2}\right)^{3/2} - \frac{1}{32}\frac{m^4}{T^4}\log\frac{m^2}{a_bT^2} + \cdots\\
J_\mathrm{F}(m^2/T^2) &\approx \frac{7\pi^4}{360} + \frac{\pi^2}{24}\frac{m^2}{T^2} - \frac{1}{32}\frac{m^4}{T^4}\log\frac{m^2}{a_fT^2} + \cdots
\end{align}
Since $T_t > T_{0 SM}$, the destabilization temperature, the temperature scale for the phase transition is much larger than any of the particle masses; a high temperature expansion will be a good approximation in this regime.

In terms of the definitions given in Section \ref{sec:backgrnd} we have
\begin{align}
\lambda(T) &= \frac{\lambda_{SM}(T)}{4}\\
a(T) &= ET\\
b(T) &= D(T^2 - T_{0 SM}^2)\\
\delta(T) \equiv \frac{8\lambda b}{a^2} &=
\frac{2\lambda_{SM}(T)D(T^2 - T_{0 SM}^2)}{E^2T^2} =
\frac{2D}{E^2}\left[1 - \left(\frac{T_{0 SM}}{T}\right)^2\right]\lambda_{SM}(T).
\end{align}

We now consider an effective potential of the general form of the SM case,
\begin{equation}\label{eq:genpot}
V_{eff}(\phi,T) = \frac{\lambda(T)}{4}\phi^4 - (ET - e)\phi^3 + D(T^2 - T_0^2)\phi^2,
\end{equation}
where there is a new parameter, $e$, motivated by gauge singlets (see Sec.~\ref{sec:singlet}).  $e = 0$ in the SM.  The potential minima (first derivative with respect to $\phi$ is $0$) located at
\begin{equation}\label{eq:genphic}
\phi = 0, \frac{3(ET - e) \pm \sqrt{9(e - ET)^2 - 8D\lambda(T)(T^2 - T_0^2)}}{2\lambda(T)}.
\end{equation}
The parameters in the Euclidean action are (using a tilde here to denote the $\lambda$ of the formulae in Sec.~\ref{sec:se3}, otherwise we mean the $\lambda$ of our general effective potential above):
\begin{align}
\tilde{\lambda}(T) &= \frac{\lambda(T)}{4}\\
a(T) &= ET - e\\
b(T) &= D(T^2 - T_0^2)\\
\delta(T) \equiv \frac{8\tilde{\lambda} b}{a^2} &= \frac{2D(T^2 - T_0^2)\lambda(T)}{(ET - e)^2}.
\end{align}
The critical temperature is (the sign of the square root is chosen so that the temperature is positive in the SM case):
\begin{equation}\label{eq:gentc}
T_c = \frac{eE-\sqrt{D\lambda(T_c)(e^2 + (D\lambda(T_c) - E^2)T_0^2)}}{E^2 - D\lambda(T_c)}
\end{equation}
From here on the temperature dependence of $\lambda$ will be dropped, taking it as a free parameter in the theory.  Alternatively, $\lambda(T) \approx \lambda(T_0)$  or $\lambda(T) \approx \lambda(T_c)$, since $\lambda$ is slowly varying with temperature (logarithmic corrections).

For calculating the tunneling temperature, we note the general behavior of $S_{E3}/T$: it decreases rapidly as the temperature is lowered, from a singularity at $T=T_c$.  Physically, this is because the tunneling rate starts at zero when the minima are degenerate, and increases rapidly as the new minima becomes the global one.  In the approximation, this can be seen by looking at the term $(2 - \delta)^{-2}$.  At $T_c$, $\delta(T_c) = 2$, for this general effective potential.  To approximate a solution for $T_t$ we expand near $T_c$: $T \rightarrow T_c(1 - \epsilon)$.  $\epsilon$ will be very small for $T_t$, as the sharp peak ensures that $S_{E3}/T = 140$ very close to $T_c$.  Expanding $\delta$ to lowest non-vanishing order in $\epsilon$, about $\epsilon = 0$,
\begin{equation}
\delta \approx 2 + \frac{4D\lambda(ET_0^2 - eT_c)T_c}{(e - ET_c)^3}\epsilon \equiv 2 + F\epsilon,
\end{equation}
where the first term is due to substituting in \eqref{eq:gentc}.  As $\epsilon \rightarrow 0$, we recover that $\delta \rightarrow 2$.  Now the relevant quantity for the tunneling temperature, $S_{E3}/T$, can be approximated by again having $T \rightarrow T_c(1 - \epsilon)$ and expanding about $\epsilon = 0$.  Defining the prefactors of $S_{E3}$ as
\begin{equation}
G \equiv \frac{64\sqrt{2}\pi}{81\lambda^{3/2}},
\end{equation}
the resulting lowest order expression is
\begin{equation}\label{eq:se3tnew}
S_{E3}/T \approx \frac{2G(\beta_1 + 2\beta_2 + 4\beta_3)(ET_c - e)}{F^2T_c}\frac{1}{\epsilon^2}.
\end{equation}
As expected, there is a singularity as $\epsilon \rightarrow 0$, with the same power as the divergent piece in the original expression.  Finally, we arrive at an estimate for the tunneling temperature:
\begin{align}
\epsilon &\approx \sqrt{\frac{2G(\beta_1 + 2\beta_2 + 4\beta_3)(ET_c - e)}{140F^2T_c}} \label{eq:epsilon}\\
T_t &\approx T_c(1 - \epsilon).
\end{align}
A rough error estimate comes from comparing this value of $T_t$ to the value obtained numerically using the first approximation, \eqref{eq:se3}.  Throughout most of the parameter space that we analyze in the next section, the difference between the two values averages to be less than $0.1\%$ (most of the space is much lower even).  Some regions of $-e$, however, can have an average difference of $30\%$.

The true minimum of the potential is $\phi$ with the positive square root in \eqref{eq:genphic} which gives an expression for the minimum of the potential as a function of temperature. This yields an exact expression for $\alpha$ (from the formulae in Sec.~\ref{sec:gwp}):
\begin{align}
\alpha &= \frac{15(3e - 3ET_t - \xi)}{16\pi^2g_t\lambda^3T_t^4}\left\{9e^3+e^2(9ET_t - 6\xi) + 9E^2T_t^2(3ET_t + \xi) - 4D\lambda\left[-3ET_t(T_0^2 - 3T_t^2)\right.\right.\notag\\ &\left.\left. + (T_0^2 + T_t^2)\xi\right] + e\left[-45E^2T_t^2 + 12D\lambda(T_0^2 + T_t^2) - 3ET_t\xi + \xi^2\right]\right\},
\end{align}
where $\xi \equiv \sqrt{9(e-ET_t)^2 - 8D\lambda(T_t^2-T_0^2)}$.  $\beta/H_t$ can be computed directly from \eqref{eq:se3} and the definitions in Sec.~\ref{sec:approx}:
\begin{align}
\beta/H_t &= \frac{1024\pi D^2\sqrt{2\lambda}}{81 T_t (e-E T_t)^{10}\sqrt{D\lambda h}k^3} \left\{-32 D\lambda T_t (E T_0^2-e T_t) (ET_t -e) h^2j(8,64)\right.\notag\\
&\left.-E T_t h^2kj(8,64) + (E T_t - e)h^2k j(8,64) + T_t (e T_t - E T_0^2)hkj(8,64)\right.\notag\\
&\left.+ 4T_t (E T_0^2 - e T_t) (e - ET_t)\left[(e - ET_t)^2 - 4 D\lambda h\right]h j(16,192)\right\},
\end{align}
where
\begin{align}
h &\equiv T_t^2 - T_0^2\\
j(x,y) &\equiv (ET_t - e)^4\beta_1 + xD\lambda(ET_t - e)^2h\beta_2 + yD^2\lambda^2h^2\beta_3\\
k &\equiv (ET_t - e)^3\left(2 - \frac{8D\lambda h}{(ET_t - e)^2}\right).
\end{align}

A much simpler expression comes from using the approximation in \eqref{eq:se3tnew} (noting that $\frac{\textrm{d}}{\textrm{d}T} = -\frac{1}{T_c}\frac{\textrm{d}}{\textrm{d}\epsilon}$ near $T_t$):
\begin{equation}
\beta/H_t \approx \frac{4G(\beta_1 + 2\beta_2 + 4\beta_3)(ET_c - e)}{F^2T_c}\frac{(1 - \epsilon)}{\epsilon^3}.
\end{equation}
Again, a rough error estimate comes from comparing this approximation versus computing $\beta/H_t$ directly from \eqref{eq:se3}.  Over the parameter space we consider, the difference typically averages to be less than $10\%$.  Again, part of the $-e$ region has much larger errors, but, on average, well within an order of magnitude.  These errors in $T_t$ and $\beta/H_t$ might potentially affect the peak frequency and overall power, to an extent that quantitatively depends on the size of these errors.  It is difficult to estimate analytically the effect on the GW spectrum, but if the errors are not very large, they are possibly only important in borderline detection cases.

Using the above approximation for the tunneling temperature provides an approximation for the GW  parameters for any theory with a SM-like effective potential.  The GW spectrum is then fully specified, using the equations at the end of Sec.~\ref{sec:gwp}.

\subsection{Parameter Constraints}
We need to enforce that the potential of \eqref{eq:genpot} (where we will work at $T = 0$ here) correctly describes electroweak symmetry breaking.  Note that the potential is typically not a tree level potential, even at $T = 0$ (in the SM there are still the one-loop effects), and using the SM parameters will give differing results (by a few percent) from tree level calculations.  The first condition is that $\phi$ at the electroweak breaking minimum, \eqref{eq:genphic} at $T=0$ with the positive root, is the usual vev of the Higgs field:
\begin{equation}\label{eq:vev}
<\phi> = \frac{-3e + \sqrt{9e^2 + 8D\lambda T_0^2}}{2\lambda} = v \approx 246\textrm{ GeV},
\end{equation}
where $\lambda$ is also at $T = 0$.  This point must be a stable minimum, hence
\begin{equation}\label{eq:stab}
\frac{\partial^2V_{eff}(\phi,T=0)}{\partial\phi^2}\bigg|_{\phi = v} = \frac{1}{2\lambda}\left(9e^2 + 8D\lambda T_0^2 - 3e\sqrt{9e^2 + 8D\lambda T_0^2}\right) > 0.
\end{equation}
Finally, we will also restrict the parameters to have a Higgs mass above current limits:
\begin{equation}\label{eq:mh}
m_h^2 = \frac{\partial^2V_{eff}(\phi,T=0)}{\partial\phi^2}\bigg|_{\phi = v} > (114 \mathrm{ GeV})^2
\end{equation}

The first constraint, \eqref{eq:vev}, can be solved to give an expression for $T_0$:
\begin{equation}
T_0^2 = \frac{v(3e + \lambda v)}{2D}
\end{equation}
which is rather similar to the SM form.  Requiring that $T_0$ and $v$ are greater than $0$, along with the original constraints on the coefficients of the potential ($\lambda, D, (ET - e) > 0$, see Section \ref{sec:se3} and \cite{se3approx}), satisfies the second constraint, \eqref{eq:stab}, if $e > 0$ or if $e < 0$ and $\lambda \neq -\frac{3e}{2v}$.  Finally, we can use the the last equation for $m_h$, \eqref{eq:mh}, to solve for $\lambda$.  In order to satisfy all the constraints we are left with the following solutions:
\begin{equation}
\lambda = 
\begin{cases}
\frac{m_h^2-3ev}{2v^2},\quad\frac{1}{4v^2}(m_h^2-9ev-\sqrt{m_h^4 - 18evm_h^2 + 9e^2v^2}),\\ \quad\quad\text{if $\left(\frac{m_h^2 - 2v^2}{3v}, \frac{-1}{6}\left(3v + \sqrt{4m_h^2 + v^2}\right)\right) < e < 0$,}\\
\frac{m_h^2-3ev}{2v^2}, \qquad\text{if $0 \le e < \frac{m_h^2}{3v}$.}
\end{cases}
\end{equation}
For the SM case of $e = 0$ we have the usual relation of $m_h^2 = 2\lambda v^2$.  In order for the theory to be perturbative we need $\lambda < 1$, so for the SM we use
\begin{equation}
0.11 \le \lambda < 1,
\end{equation}
where the lower bound is from the Higgs mass limit.  This roughly corresponds to
\begin{equation}
115\textrm{ GeV} \le m_h < 348\textrm{ GeV.}
\end{equation}
When $e \ne 0$ the value of $e$ and $m_h$ in the range given above set the value of $\lambda$.  When $e < 0$ both solutions for $\lambda$ are used, which gives the greatest range of values for both $\lambda$ and $e$.

Other constraints follow from requiring $T_c$ to be real and positive, which can be deduced simply from \eqref{eq:gentc}.  For instance, for $T_c$ to be real,
\begin{equation}
e^2 \ge (E^2 - D\lambda)T_0^2,
\end{equation}
while having $T_c > 0$ can reduce to a simple constraint depending on which parameters are being varied together, as well as the sign of the square-root.

It is also important to note divergent features of the above expressions.  First, in \eqref{eq:epsilon}, $\epsilon$ will be complex if $e > ET_c > 0$, since the other terms are all positive ($D$ and $E$ are positive in general, as they depend only on masses-squared).  In this case $T_t$ and the GW parameters will also be complex, so we limit
\begin{equation}
e < ET_c.
\end{equation}
Since this arises in our approximation, it is possible that higher order terms will resolve this behavior.  However, if $e \ge ET_c$, then the sign of the cubic term in the potential changes (becoming positive).  There would no longer be any potential energy barrier, and the phase transition could not be first-order.  Therefore, it is consistent for this analysis to limit $e$ by the inequality above.

In several of the equations in Section \ref{sec:approx} the term $E^2 - D\lambda$ appears.   Occurring in the numerator of eq.~\eqref{eq:gentc} this appears in the reality constraint above.  Appearing in the denominator, however, when it equals zero it is a pole in the (exact) expression for $T_c$.  This causes $\epsilon$ to diverge, and thus $T_t$ will pass through zero.  This carries over to a pole in $\alpha$.  Clearly, as $\epsilon$ approaches one, our approximation will break down.  However, fine-tuning to be near this pole could provide models with a very strong phase transition (large $\alpha$).  Below, we only consider values of $E^2$ up to $90\%$ of $D\lambda$.

\section{The Parameter Space}\label{sec:param}

Now that we have general expressions for the GW parameters in terms of the coefficients of our potential, we can look at the parameter space.  Throughout this analysis the parameters we are not varying are set to their SM values (with $\lambda_{SM} = 0.3$, and $\lambda_{SM}(T)\big|_{T=T_0}$ when not varied).  Also, we set $g_t = 100$ throughout.  This can vary greatly between models, but here we are looking at the parameter space in general, without assumptions on what theory is being used.

\begin{figure}
\begin{center}
\includegraphics[width=14cm,clip]{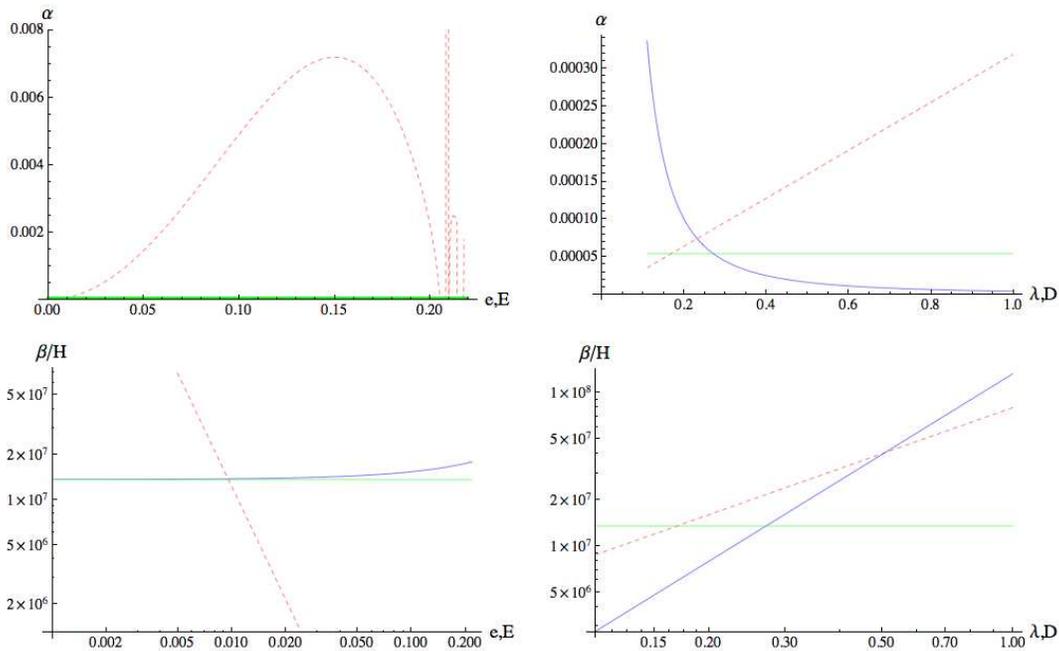}
\caption{In each plot the labels for the x-axis denotes the varied parameter in blue and red (dashed), respectively.  The green line is the (constant) SM value.  The ranges for the varied parameters were chosen arbitrarily, to show the overall behavior.  $e$ is in GeV, while the other parameters are dimensionless.  Note the pole as $E$ is varied, as described.  On the left-hand side, the blue line (varying $e$) closely follows the green one (SM value), for the range shown.}
\end{center}
\end{figure}

\begin{figure}
\begin{center}
\includegraphics[width=14cm,clip]{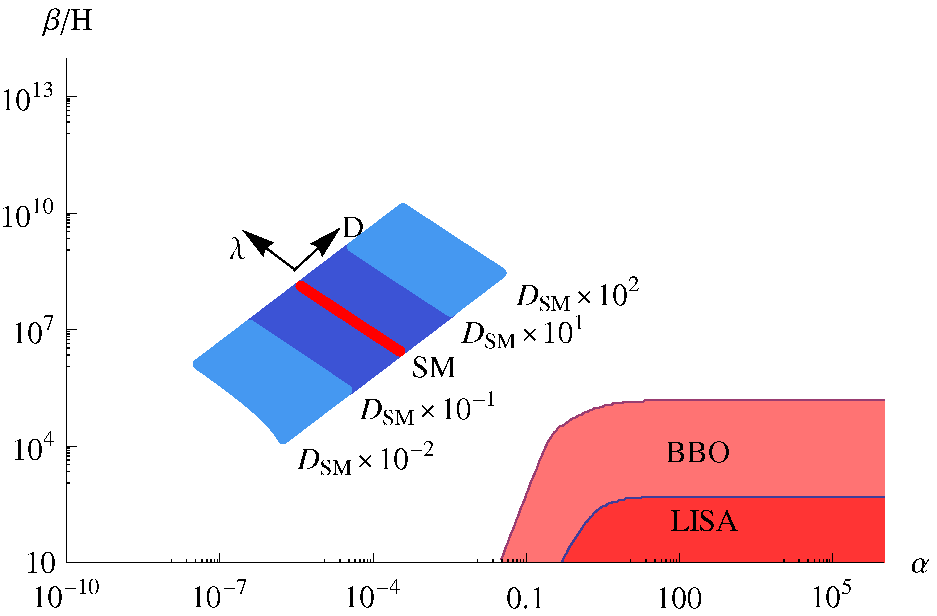}
\caption{A plot of the $\alpha-\beta/H$ plane with $\lambda$ and $D$ varying.  The red line is the SM with $\lambda$ in the allowed range.  The darker blue region is an order of magnitude greater and less than the SM value for $D$.  The lighter blue region is a further order of magnitude greater and less.  BBO and LISA rough sensitivity regions are shown as the lighter and darker red shading, respectively.}
\end{center}
\end{figure}

\begin{figure}
\begin{center}
\includegraphics[width=14cm,clip]{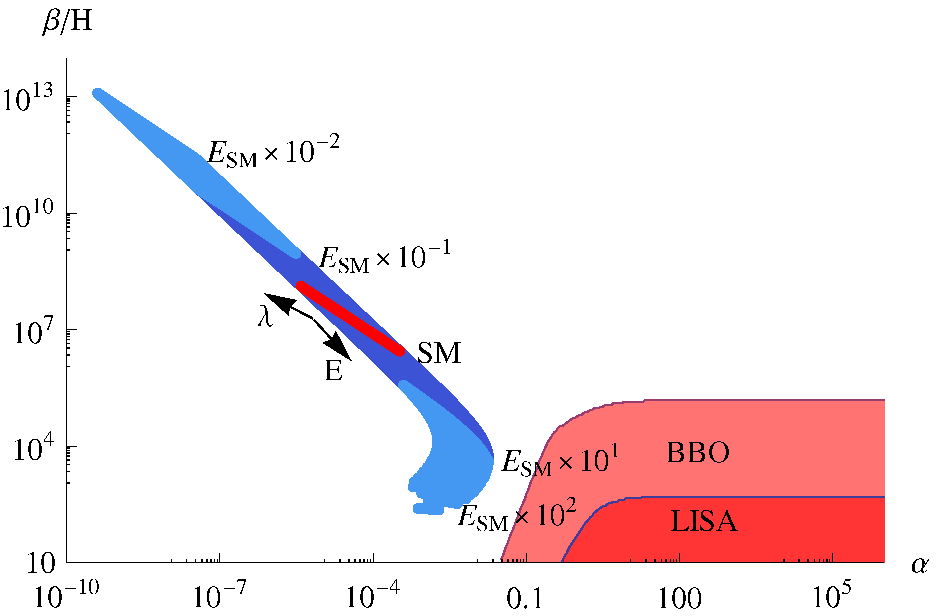}
\caption{A plot of the $\alpha-\beta/H$ plane with $\lambda$ and $E$ varying.  The red line is the SM with $\lambda$ in the allowed range.  The darker blue region is an order of magnitude greater and less than the SM value for $E$.  The lighter blue region is a further order of magnitude greater and less.   BBO and LISA rough sensitivity regions are shown as the lighter and darker red shading, respectively.}
\end{center}
\end{figure}

\begin{figure}
\begin{center}
\includegraphics[width=15cm,clip]{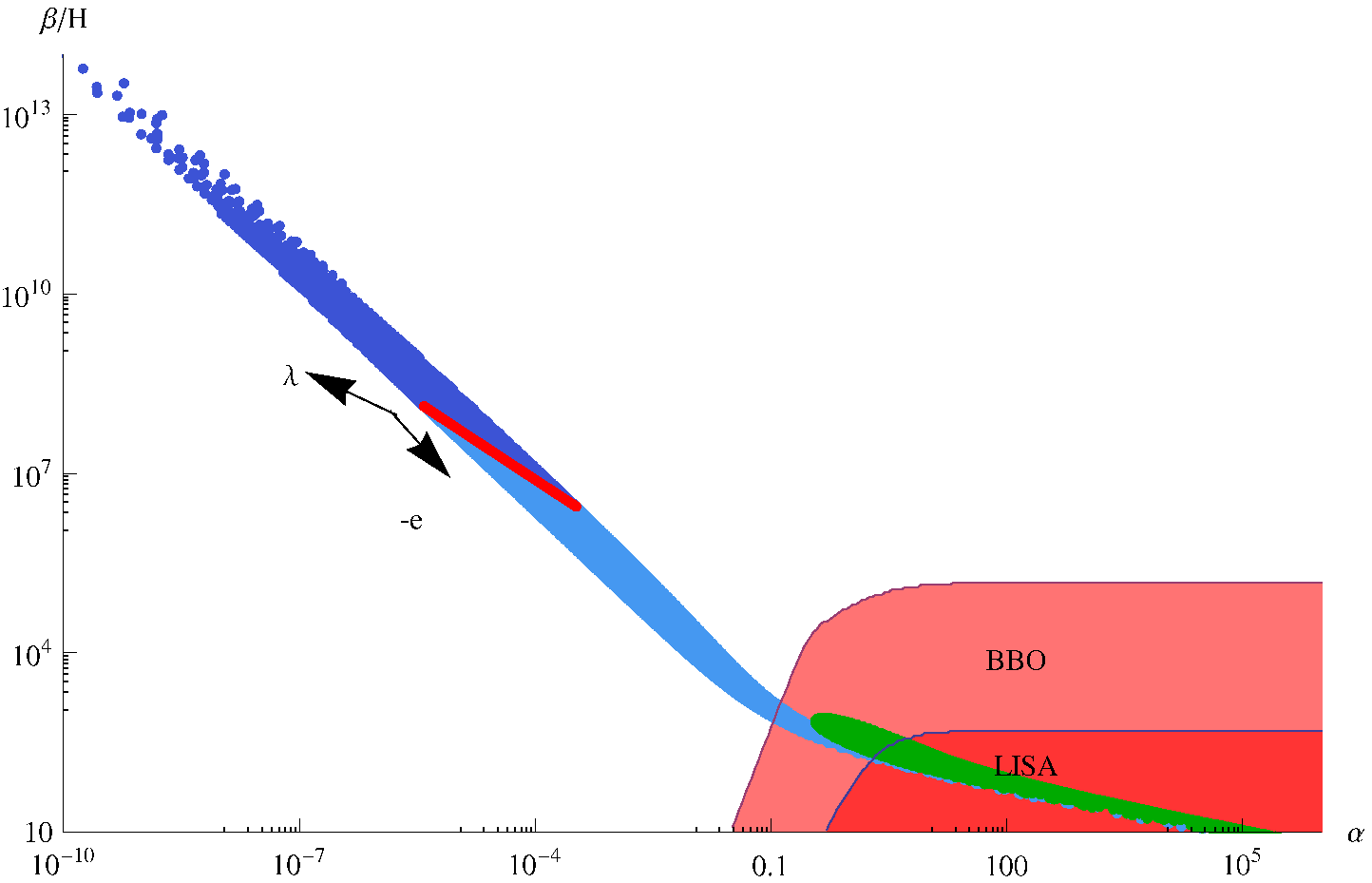}
\caption{A plot of the $\alpha-\beta/H$ plane with $\lambda$ and $e$ varying.  The dark blue region is for $e > 0$, while the light blue shading indicates the first solution for $\lambda$ when $e < 0$, and the green the second solution.   BBO and LISA rough sensitivity regions are shown as the lighter and darker red shading, respectively.  The tail of the green region, extending to very large $\alpha$, may have errors in the calculation of $\beta/H_t$ due to the approximate formula.  However, the points remain in the LISA region.}
\end{center}
\end{figure}

In the plots of the $\alpha-\beta/H_t$ plane both $\lambda$ and one other parameter are varied.  $\lambda$ ranges as described above.  The other varied parameter ranges from an order of magnitude less than its SM value to an order of magnitude larger for the darker blue region, and a further order of magnitude larger and smaller for the lighter blue region, taking into account the above constraints.  Note that for negative $e$, we also need to make sure that $\epsilon$ is real and less than one, as this is the only parameter (in the ranges we considered) that could cause these problems for $\epsilon$.  The red line represents the SM with only $\lambda$ varied.  $e$ is less than $ET_c$ or $m_h^2/3v$, whichever is smallest.  LISA and BBO sensitivity regions were computed using rough estimates of the sensitivity curves, as in \cite{gwbubble,Buonanno:2004tp,Grojean:2006bp,Caprini:2009yp}, and finding the region in the $\alpha$-$\beta/H$ plane where the computed GW spectrum is above the sensitivity.  The LISA spectrum used is the approximate instrument strain sensitivity, while the BBO spectrum is the correlated extension (BBO Corr), which uses data analysis of two LISA-like detectors to improve sensitivity.  LIGO is not sensitive in this region to GWs.  The sensitivity regions shown flatten (do not reach into higher $\beta$) for large $\alpha$, as the LISA and BBO sensitivity drops considerably outside of their most sensitive frequency.  Thus, even with increasing $\alpha$, larger $\beta$ means a greater peak frequency, pushing it out of the experimental range.

We note that very small $v_b$, due to viscosity of the plasma, could potentially affect detection with LISA or BBO, due to a change in the peak energy density and the peak frequency of the GWs.  A small $v_b = 0.05$, as in \cite{vb}, does not change the overall overlap of the experimental sensitivities with the parameter space in the figures (using this $v_b$ in calculating the experimental sensitivity).  However, this could be a large effect for borderline cases.  Without specifying the specifics of a theory, it is difficult to explore the issue of calculating $v_b$ precisely.  Furthermore, our analysis considers only detonations, where $v_b$ is at least ten times greater.  Small $v_b$ requires considering the case of deflagrations in detail, and is outside the scope of this work (for recent work, see \cite{Megevand:2008mg,Caprini:2007xq}).  This is another direction to be explored not only in the context of GW detection, but also for baryogenesis, as $v_b$ affects the length of the phase transition.

Although varying $D$ or $E$ does not greatly enhance $\alpha$, their effect is noticeable.  Both also tend to have large values of $\beta$, but cover a very large range.  However, $D$, at least for this range of variation, is not close to the sensitivity regions of LISA or BBO.  $E$ has regions which are much closer to LISA or BBO, but we do not predict any overlap.  $e < 0$, on the other hand, can greatly increase $\alpha$, as well as have a much smaller $\beta$.  Here there is the greatest parameter space which may be observable.  There is considerable overlap between the various potential parameters and the corresponding values of the GW parameters, and thus a particular GW spectrum does not generally point to a value of just one of these potential parameters.

\section{Models}\label{sec:models}

In this section we investigate a few models which naturally fit into the effective potential analyzed above.  Of primary interest are models with gauge singlets, which we argue provide a very good motivation for the parameter $e$.  Since this parameter was shown to have the largest effect in producing a strongly first-order phase transition, these models could predict an observable GW signal in the near future.  There are many models with singlets, and in general it is not possible to derive an analytic expression for the effective potential.  We will look at the simplest such model, the SM with the addition of one (real) gauge singlet field.

\subsection{SM + Gauge Singlet}\label{sec:singlet}
While the motivation for the overall form of the effective potential we consider is due to the SM, the additional term, $e$, is motivated by gauge singlets.  Gauge singlets are present in many models, such as the Next-to-MSSM (NMSSM).  Singlets in the context of the electroweak phase transition and baryogenesis have been studied for some time (e.g.~\cite{singlet1,singlet2,singlet3}, and more recently, e.g.~\cite{ahriche,profumosinglet,darkside}), including the effect of trilinear couplings driving a strongly first-order phase transition without even considering the thermal term (see, for instance, \cite{cubic1,cubic2}).  Recently, \cite{darkside} has also explored the connection of singlets with possible explanations for recent astrophysical signals.  In this section we will give a brief decoupling (in the sense of taking the singlet to be much heavier than the Higgs) analysis, motivating the inclusion of $e$ in the effective potential, and showing that these classes of models naturally fit into our framework above.

The procedure we will follow is to use the tree-level, zero-temperature mass eigenstates as a basis, taking the mostly singlet state to be much heavier.  In addition, we will want the mixing of the states to be very small.  The goal is to show that there is a tree-level, cubic, self-coupling term of the lighter (mostly Higgs) state.  We will consider two different limiting cases (in terms of an expansion parameter, defined below) and show that we can always have a very small mixing angle, a very heavy singlet state, and a light Higgs state with sufficiently large cubic coupling.  If we consider SM values for all other parameters of the effective potential, than we need $e \lesssim -20$ GeV (see previous section) for a strongly first-order phase transition.

Following the notation of \cite{profumosinglet}, the tree level scalar potential of the SM Higgs sector with an additional gauge singlet field is the sum of
\begin{align}
V_{SM} &= -\mu^2\left(H^\dagger H\right) + \lambda\left(H^\dagger H\right)^2\\
V_{HS} &= \frac{a_1}{2}\left(H^\dagger H\right)S + \frac{a_2}{2}\left(H^\dagger H\right)^2S^2\\
V_S &= \frac{b_2}{2}S^2 + \frac{b_3}{3}S^3 + \frac{b_4}{4}S^4
\end{align}
with $H$ the usual SM $SU(2)_L$ scalar doublet and $S$ the gauge singlet (real) scalar field.  The vevs are defined to be $v_0/\sqrt{2}$ for $H$ and $x_0$ for $S$.    The fields for the fluctuations about the vevs are $h$ and $s$, defined as $H = (v_0 + h)/\sqrt{2}$ and $S = x_0 + s$.  For simplicity, we'll now take $H$ to be real.  The minimization conditions (for both fields, with $x_0 \neq 0$) can be used to eliminate the mass parameters:
\begin{align}
\mu^2 &= \lambda v_0^2 + (a_1 + a_2x_0)\frac{x_0}{2}\\
b_2 &= -b_3x_0 - b_4x_0^2 - \frac{a_1v_0^2}{4x_0} - \frac{a_2v_0^2}{2}.
\end{align}
While for $x_0 = 0$ the conditions enforce $a_1 = 0$ and $\mu^2 = \lambda_Sv_0^2$, as in the SM.

The mass matrix has the following elements:
\begin{align}
\mu_h^2 \equiv \frac{\partial^2 V}{\partial h^2} &= 2\lambda v_0^2\\
\mu_s^2 \equiv \frac{\partial^2 V}{\partial s^2} &= b_3x_0 + 2b_4x_0^2 - \frac{a_1v_0^2}{4x_0}\\
\mu_{hs}^2 \equiv \frac{\partial^2 V}{\partial h\partial s} &= (a_1 + 2a_2x_0)v_0.
\end{align}
The mass eigenstates are defined as
\begin{align}
h_1 &= (\sin\theta) s + (\cos\theta) h\\
h_2 &= (\cos\theta) s - (\sin\theta) h
\end{align}
with the mixing angle $\theta$ as
\begin{equation}
\tan\theta = \frac{y}{1 + \sqrt{1 + y^2}}, \qquad \textrm{where } y \equiv \frac{\mu_{hs}^2}{\mu_h^2 - \mu_s^2}.
\end{equation}
Here, $|\cos\theta| > 1/\sqrt{2}$, and so $h_1$ is the state with the largest $SU(2)$-like component (and $h_2$ has the largest singlet component).  The terms singlet state, $h_2$, and heavier state will be used interchangeably (and likewise for the Higgs state).  Inverting the above states, we have the original fields, expanded about the minimum, in terms of these mass eigenstates:
\begin{align}\label{eq:states}
H &= v_0 + (\cos\theta) h_1 - (\sin\theta) h_2\\
S &= x_0+ (\sin\theta) h_1 + (\cos\theta) h_2.
\end{align}
The mass eigenvalues are
\begin{equation}
m_{1,2}^2 = \frac{\mu_h^2 + \mu_s^2}{2} \pm \frac{\mu_h^2 - \mu_s^2}{2}\sqrt{1 + y^2},
\end{equation}
with the upper (lower) sign for $m_1$ ($m_2$).

We consider the decoupling limit where the singlet is very heavy (i.e.~the state $h_2$) and study the cubic term of the effective $h_1$ potential.  The lighter state, $h_1$, should roughly be in the mass range allowed for the SM Higgs.  Although we allow mixing between the states (so the singlet will not be completely physically decoupled), we want to keep the mixing angle small; interactions between $h_1$ and $h_2$ will be highly suppressed after integrating $h_2$ out, by both the small mixing angle and large mass scale of $h_2$.  However, since higher dimensional operators are generated, this could contribute to enhancing $E$ (and thus $\alpha$), as discussed in \cite{dim6}.  We will also consider $x_0$ an essentially free parameter, determined by physics at the higher, singlet scale.  We will consider $x_0$ both much larger or smaller than $v_0$, and use the ratio as an expansion parameter.

Since we consider $x_0$ set by other dynamics, $S$ will be expanded as in eq.~\eqref{eq:states}, while we will drop the $v_0$ in expanding $H$\footnote{If we include $v_0$, the only change is the addition of the expected cubic coupling of the shifted Higgs field in the SM.}.  The reason for this is that we want to consider these results as leading terms in the finite-temperature effective potential; the proper degree of freedom before the phase transition is the unshifted field (there is no vev yet).  Clearly, the mass eigenstates above are for the tree-level, zero-temperature, shifted fields.  However, we will consider this as simply a change of basis; the coefficients of the expansion are just the proper values for the mass eigenstate basis at tree level (and zero temperature).

Considering the limit of $x_0 \gg v_0$, we expand to lowest order in the small parameter $u \equiv v_0/x_0$.  We have that $\cos\theta \approx 1$ and $\sin\theta \approx -\frac{a_2}{2b_4}u$.  Also, to lowest order in $u$,
\begin{align}
m_1^2 &\approx \frac{(4b_4\lambda - a_2^2)v_0^2}{2b_4}\\
m_2^2 &\approx \frac{2b_4v_0^2}{u^2}.
\end{align}
Note that as we take $u$ very small, the heavy mass becomes arbitrarily large.  Writing the fields in the potential in terms of the mass eigenstates and working to lowest order in $u$, the coefficient of $h_1^3$ is given by
\begin{equation}
-\frac{a_2^2v_0}{2b_4} + \frac{a_2}{2b_4}\left(\frac{a_2b_3}{2b_4} - a_1\right)u.
\end{equation}
Some fine tuning will be needed to have appropriate masses and cubic coupling.  The values, for example, of $a_2 = 0.35, b_4 = 0.3, \lambda = 0.4$, has a light mass of about $190$ GeV and a cubic coupling of about $-50$ GeV (in the limit of $u \rightarrow 0$).  The massive parameter $a_1$ could also be tuned as $1/u$, for instance.  In this case the series expansions also change and there are large regions of parameter space where all the constraints can be met.


In the small $x_0$ limit, to lowest order in $w \equiv x_0/v_0$,
\begin{align}
\theta &\approx 2w\\
m_1^2 &\approx 4\lambda v_0^2 + a_1v_0w\\
m_2^2 &\approx -\frac{a_1v_0}{4w}.
\end{align}
Again, the heavy state mass is arbitrarily large as we take $w$ very small.  The cubic coefficient of $h_1$, in this limit, is simply $a_1w$.  In this case, we can have $m_2$ large and a sufficient cubic coupling by having $a_1$ negative and (possibly unnaturally) large.  However, the second term in the expression for $m_1^2$ can constrain $a_1$ (or require tuning with $\lambda$).  For instance, the light state has mass about $177$ GeV when $\lambda = 0.3$ and $a_1 = -20/w$ (the cubic coupling is $-20$ GeV), with $w$ arbitrary.  Note that these expansions (to lowest order) do not change for $a_1 = a/w$.

In both limiting cases then, a cubic coupling in the effective potential of the light state appears.  This coupling can be of the right size to drive a strongly first-order phase transition, while still having a small mixing angle between the states, and appropriate heavy/light masses for the fields.  It is also possible (for instance, \cite{profumosinglet}) to approximate the two-field effective potential in the same form as we analyzed (with the field being the Higgs, not the singlet).  Viewed this way, the coefficients are more general functions of the singlet field.  In general, however, this potential still must be numerically analyzed, due to the dynamics of the Higgs-singlet interactions.  The decoupling analysis shown above is simply one limit of a general analysis.  It is meant to motivate $e$ and show that, even in this extreme limit, a singlet can have a large effect on the phase transition\footnote{The singlet may also have other effects on the overall potential, including from its own finite temperature potential.  A full analysis, of course, must account for effects besides generating $e$.}.

\subsection{SM + $SU(2)_L$ Triplet}
An example of a model where the quartic coupling in the tree-level effective potential is suppressed for a fixed value of the SM-like Higgs is given by models with an additional $SU(2)_L$ triplet in the scalar sector.  First considered in \cite{tripletfirst}, this extension to the SM Higgs sector has several significant phenomenological implications, including a dark matter candidate \cite{Cirelli:2005uq, Cirelli:2007xd} and providing a natural framework for a Type II seesaw mechanism for generating non-vanishing neutrino masses \cite{triplet}.  We postpone an exhaustive study of this scenario to a future paper \cite{profumopavel}, but we point out here this scenario as one where the results of the present analysis apply.

Indicating with $\delta$ the neutral component of the $SU(2)_L$ triplet $\Delta$, the tree level neutral CP-even part of the Higgs potential, neglecting interaction terms with two or four $\Delta$'s, reads:
\begin{equation}
V(H,\delta) = -m_H^2H^2 + \frac{\lambda_{SM}}{4}H^4 + M_\Delta^2\delta^2 - 2\mu\delta H^2,
\end{equation} 
where $M_\Delta$ is the mass term associated to $\Delta$ and the last term stems from the term, in the scalar potential,
\begin{equation}
\mu H^T\ i\sigma_2\Delta^\dagger H+{\rm h.c.}
\end{equation}
(see e.g.~\cite{triplet}).  Imposing the minimization condition to the tree level potential and singling out the field-space trajectory along which the minimum of the (tree-level) potential is found allows us to express $\delta$ as a function of $H$:
\begin{equation}
\frac{\partial V}{\partial \delta}=0\quad \rightarrow \quad \delta=\frac{\mu}{M_\Delta^2}H^2.
\end{equation}


Along the locus of minima in the $\delta$ direction, the effective potential is now a function of a single field, $H$:
\begin{equation}
V_{min}(H) = -m_H^2H^2 + \left(\frac{\lambda_{SM}}{4} - \frac{\mu^2}{M_\Delta^2}\right)H^4.
\end{equation}
This is essentially the same as in the SM case, but with the parameter change
\begin{equation}
\frac{\lambda_{SM}}{4} \longrightarrow \frac{\lambda_{SM}}{4} - \frac{\mu^2}{M_\Delta^2} \equiv \lambda_T.
\end{equation}

This model can therefore be cast into the form of the effective potential of the SM we consider in this study, with the above change to the effective quartic coupling $\lambda$.  The extra triplet scalar therefore automatically enhances the strength of the phase transition, since, for a given SM Higgs mass and at a given tri-scalar coupling $E$, a smaller $\lambda$ correspond to larger values of $\alpha$ (which we can think of as the ``strength'' of the phase transition).  

\subsection{Other Models}
Here we will only briefly mention a few other models which could be analyzed in our framework.  First is the analysis of \cite{higherdim}.  The authors study baryogenesis using the MSSM to generate a low energy effective theory, and in particular also study the strength of the electroweak phase transition.  Given certain constraints on the parameters, the potential minimization reduces to the one-dimensional case, and the phase transition strength (i.e.~the value of $\alpha$) can be greatly enhanced.  It is enhanced through an increase of the parameter $E$, which can be about an order of magnitude larger than in the SM.  Given SM values for the other parameters in the effective potential, this alone, for a light Higgs, drives $\alpha$ to be larger than one.  The new expressions for $E$ can also be used to investigate more closely the parameter space that produces large $\alpha$.

In a similar vein, \cite{dim6} considers dimension-six Higgs operators, arising, for instance, from integrating out a heavy singlet.  This can also produce a first-order phase transition.  Another effect of this operator is to alter the Higgs self-couplings from the SM.  The cubic and quartic self-couplings are altered, and this in turn would alter the couplings appearing in the effective potential.  Again, this is easily incorporated in our analysis.

The ``topflavor'' model also has a phase transition which fits into our framework.  In this model there are separate $SU(2)$'s for the third versus other generations.  In \cite{topflavor}, the earlier phase transition, from $SU(2)_1 \times SU(2)_2$ to the SM $SU(2)_L \times U(1)_Y$, was analyzed in the context of baryogenesis.  This phase transition has a scalar field as the order parameter, which has a quartic (tree-level) potential.  The one-loop, finite-temperature, effective potential can be derived, and it has the same functional structure as the SM.  The high temperature expansion will then be the same form as the general effective potential we analyzed.  By using the constants of the topflavor model, which has a strongly first order phase transition, the GW parameters can be found through the above results.  Note that in this case some of the constraints on the parameters are not applicable, since they refer to specific electroweak constraints.  Baryogenesis in this model requires that the electroweak phase transition is \emph{not} strongly first order, and therefore any GW produced comes from this earlier phase transition.

Finally, we note that the form of the potential we used is really quite general.  The potential mimics the SM form, which we might reasonably expect to be a good approximation to any low energy, effective theory.  Also, besides $e$, the form of the potential and the constants are all derived from the high-temperature expansion of the general one-loop, finite-temperature, effective potential of a theory with gauge bosons and fermions coupled to a scalar field.  As we have shown above, the addition of gauge singlets motivates the inclusion of the additional parameter.  Thus we expect this potential to arise very generically for any field theory model.  The obvious exception is for multiple fields controlling the phase transition (so that all but one cannot simply be integrated out).  In that case, it may still be possible to use this potential, by approximating field configurations or by taking certain limits.

\section{Conclusions}

Upcoming experiments may soon observe the first GW signals, including those from early universe processes.  Such experiments will deliver the very exciting prospect of observing GWs from the electroweak phase transition, if it is strongly first-order.  While there have been many studies on this topic, they largely involve numerical computations of the phase transition temperature and GW parameters.  In this work we have presented an analysis of a generic effective potential which is motivated by the SM Higgs potential and an additional contribution from a gauge singlet (or other possible origins) in the form of a temperature-independent cubic coupling.

By approximating the tunneling temperature to be very close to the critical temperature (based on the general form of the action), we have derived an expression for the tunneling temperature based on the parameters of the effective potential.  Using this result, the GW parameters also have expressions depending on the parameters of the potential.

Once these expressions were found, the parameter space of the potential was explored starting from the SM values.  While all the parameters can have noticeable effects, $e$, motivated from singlet models, easily has the greatest effect.  Following this, we showed how this parameter arises from a decoupling analysis of a simple SM plus singlet model.

Other models can also be analyzed with our effective potential.  The addition of an additional $SU(2)$ triplet affects $\lambda$, while higher dimensional operators can also affect $E$.  Increasing $E$ by an order of magnitude from its SM value could also produce a strongly first-order phase transition.  The topflavor model also has an effective potential of the same form as the SM for its earlier phase transition, which can also be analyzed through this work.  In this case, the GW signal would not be from the (later) electroweak phase transition.

Finally, although the effective potential studied has clear origins in the SM and singlet extensions, the potential is indeed very general.  A low energy, effective theory attempting to model the electroweak phase transition will likely closely model this form of the potential.  Additionally, the potential is a high-temperature approximation to a one-loop, finite temperature, effective potential of a scalar field with gauge bosons, fermions, and singlets.  So in this sense its form is also generic and expected to be common in a phase transition with a scalar field order parameter.

The potential applications of the results presented include applying it to other models for the electroweak (or other) phase transition.  By approximating or otherwise finding a way to put an effective potential in the form we analyzed, the tunneling temperature and GW spectrum is now easily obtained through the above formulae.  
Besides being significant itself, the detection of a stochastic background of GWs may provide insight into the specifics of models of the electroweak phase transition, and the fundamental mechanism underlying the generation of the baryon asymmetry of the universe.

\begin{acknowledgments}
We gratefully acknowledge Lam Hui for useful discussions and feedback on
this manuscript, and especially for first bringing the idea elaborated on
in this paper to our attention.  We also acknowledge the many helpful comments and  corrections (especially regarding turbulence) from the anonymous JCAP referee.  S.P.~is partly supported by an Outstanding Junior Investigator Award from the US Department of Energy (DoE), Office of Science, High Energy Physics, and by DoE Contract DEFG02-04ER41268 and NSF Grant PHY-0757911.  J.K.~was partially supported by a Doctoral Student Sabbatical Fellowship and an Orals Fellowship from UCSC.
\end{acknowledgments}

\renewcommand{\bibsection}
{\section*{References}}
\bibliographystyle{JHEP}
\bibliography{gw_ewpt_refs}
\end{document}